\documentstyle[preprint,eqsecnum,aps]{revtex}
\begin{document}
\begin{titlepage}
\draft
\title{A LATTICE SIMULATION OF THE\\
       SU(2) VACUUM STRUCTURE \footnotemark[2] }
\footnotetext[2]{Work supported in part by the
                 U.S. Department of Energy (DOE)
                 under contract DE-FG02-91ER40676. }
\author{A.\ R.\ Levi}
\address{Department of Physics,
        Boston University, \\
        590 Commonwealth Ave.,
        Boston, MA 02215, U.S.A.\\
        and}
\author{J.\ Polonyi \footnotemark[1] }
\address{Universit\'e Louis Pasteur,
         Department of Physics,
         67084 Strasbourg, Cedex, France, \\
         and \ \  CRN Strasbourg,
         67037 Strasbourg, Cedex, France}
\footnotetext[1]{On leave from L. E\"otv\"os University and
                 CRIP, Budapest, Hungary.}
\maketitle
\begin{abstract}
In this article we analyze the vacuum structure of
pure SU(2) Yang-Mills
using non-perturbative techniques.
Monte Carlo simulations are performed for the
lattice gauge theory with external
sources to obtain the effective potential.
Evidence from the lattice gauge theory
indicating the presence of the
unstable mode in the effective potential is reported.

\ \\
BU-HEP-95-15

\end{abstract}
\vfill
\end{titlepage}
\eject
\widetext

\section{INTRODUCTION}

Despite considerable progress a complete
solution of non-abelian gauge theories
has yet to be found.
The ultraviolet properties of such theories have been
well analyzed, but
even in the simplest case, i.e. SU(2) without matter,
little is known about their infrared properties and
vacuum structures.
In order to gain a better understanding of these theories a
necessary first step is the study of the vacuum.
For a general review of the vacuum
structures see \cite{paris}.
In this article we will show some lattice
techniques and simulations
that can be important to reveal some aspects of the vacuum.

Several lattice Monte Carlo simulations \cite{fourD}
have been done to examine the
vacuum structures. These simulations are usually
very difficult because of the
weakness of clear signals.
The lattice simulation that we are proposing in this article
use a method that remove the zero modes, and increase
the possibilities of extracting information
of the vacuum structures.

During the last few years many authors \cite{threeD} have also
explored the 3-dimensional case.
It is advantageous to use 3-dimensional lattices
because of the possibility to obtain cleaner data
and larger volumes.

The SU(2) case should not be fundamentally different from other
non-abelian gauge theories. In fact, there is
no reason why the QCD vacuum should differ
qualitatively from the SU(2) case. However, since the SU(2) case
is easier to implement in lattice calculation,
it was chosen for our analysis.

\section{EFFECTIVE POTENTIAL FOR NON-ABELIAN YANG-MILLS THEORIES}

A powerful method to investigate the properties of
Yang-Mills theories is to
compute the effective potential
in the background gauge \cite{dewi}.
This manifestly gauge invariant scheme is
based on the observation that
the loop expansion corresponds to
an expansion in the parameter $\hbar$ which multiplies
the entire action. Hence a shift of the fields or
a redefinition of the division of the Lagrangian
into free and interacting parts can be performed
at any finite order of the loop expansion
without violating the manifest
gauge invariance.

Let us split the gauge field into a background, ${\cal A}_\mu^b$,
and a quantized field, $\eta_\mu^b$, as
$A^b_\mu~=~{\cal A}_\mu^b~+~\eta_\mu^b $.
Therefore the effective action will also be a functional
of ${\cal A}_\mu^b$.
The gauge fixing condition is that the covariant
four-divergence of the
quantum field computed on the background vanishes,
\begin{equation}
    D_\mu\eta_\mu^b~\equiv~(\partial^\mu\delta^{bc}
    -i g {\cal A}^d_\mu (T^d)_{bc} ) \eta_\mu^c~=~0 .
\label{modigliani}
\end{equation}
The generating functional of connected Green's functions is
consequently defined as:
\begin{equation}
     e^{{i\over\hbar}\tilde W[J,{\cal A}]}~=
     ~\int~(d\eta)~
     \Delta({\cal A},\eta)~
     e^{{i\over \hbar}\{ S[{\cal A} +\eta]+(J,\eta)-
     {1\over 2\alpha} (G^a)^2 \} } ,
\label{montalcini}
\end{equation}
where $\Delta({\cal A}_\mu^b,\eta^b_\nu)$
is the Faddeev-Popov determinant,
and $G^a$ is the gauge-fixing term.
As pointed out by \cite{lehu}
several subtleties arise from the gauge fixing condition,
due to the choosing of the $\alpha$ gauge fixing parameter in the
presence of a non-trivial background. A complete
discussion of this point is beyond the scope of this article.

Following Abbot's notation \cite{dewi} ,
let us define $\tilde Q={\delta \tilde W[J,{\cal A}] / \delta J}$.
Therefore the Legendre transform of $\tilde W$ is
\begin{equation}
     \tilde\Gamma[\tilde Q,{\cal A}]~=
     ~\min_{ \{ J \} }
     ~\biggl[ \tilde W[J,{\cal A}]-(J,\tilde Q) \biggr],
\label{gadda}
\end{equation}
and the inverse relation is:
$J=-\delta \Gamma[\tilde Q,{\cal A}] / \delta \tilde Q $.

Using the background field as discussed above,
the Lagrangian for the
SU(2) Yang-Mills theory is written as:
\begin{eqnarray}
     {\cal L}(A_\mu^b={\cal A}^b_\mu+\eta^b_\mu)~=~
     &-~{1\over 4}~F_{\mu\nu}^b~F^{\mu\nu}_b~=
     ~-~{1\over 4}~F({\cal A})_{\mu\nu}^{b}~
     F({\cal A})^{\mu\nu}_b~+    \nonumber \\
     &-~{1\over 2}\eta^b_\mu~(-D^2\delta^{\mu\nu}~+~D^\mu~D^\nu
     )^{bc}~\eta_\nu^c~
     +~g~\varepsilon^{bcd}~\eta^b_\mu~F({\cal A})^{\mu\nu}_c
     \eta^d_\nu~+      \nonumber \\
     &-~g~\varepsilon^{bcd}~(D^\mu \eta^\nu)^b~
     \eta^c_\mu~\eta^d_\nu~
     -~{1\over 4}~g^2~\varepsilon^{bcd}~\varepsilon^{bef}~
     \eta^c_\mu~\eta^d_\nu~\eta^e_\rho~\eta^f_\sigma
     ~g^{\mu\rho}~g^{\nu\sigma} \ .
\end{eqnarray}

As shown in \cite{const}
there are only two possible backgrounds that yield a static
chromomagnetic field that is the interesting field configuration
to study the vacuum properties.
One is the so-called ``non-abelian background''
(see \cite{forgot} and \cite{lehu}), the other
is called ``Abelian'', and is the one that we will discuss in this article.

Without loss of generality we can always choose the chromomagnetic field
along the $z$ direction, the abelian background can be written as:
\begin{equation}
     {\cal A}_\mu^b={1\over 2}~H~\delta^{b3}~
     (\delta_{\mu 2}~x~-~\delta_{\mu 1}~y) ,
\label{allen}
\end{equation}
where $H$ is constant. We have
$F({\cal A})F({\cal A})={1\over 2} H^2$.
As discussed in \cite{mmrt} and \cite{niol}
after several manipulation the effective potential can be evaluated
using:
\begin{equation}
      -{V(H)\over \Omega T}~=~{1\over 2}H^2~
      +~{gH\over 8\pi^2}~\biggl(
      \int_{-\infty}^\infty dk_z~\sum_{n=0}^\infty~
      \sum_{s=\pm 1}~ \sqrt{\nu_{k,n,s}~}~
      -2~\int_{-\infty}^\infty dk_z~\sum_{n=0}^\infty~
      \sqrt{\tilde\nu_{k,n}~} \biggr)  ,
\label{garibaldi}
\end{equation}
where $\Omega T$ is the 4D volume, and
$\nu_{k,n,s}=(2n+1+2 S_z)gH + k_z^2 $ are the eigenvalue of
the quadratic part of the Lagrangian.
To derive this expression, the multiplicity of
each eigenvalue, the overall factor $2$ of charge degeneracy,
and the ghost contribution has been taken in account.
The last term is the contribution of the ghosts,
that is the eigenvalue of the operator $(-D^2)^{bc}$
with the same boundary conditions as for the gluon sector.
One finds $\tilde\nu_{k,n}=k^2_z+(2n+1)gH$.
To compute the expression (\ref{garibaldi})
we have to regularize using the
Salam and Strathdee method.

This expression yields to the famous Savvidy \cite{savv} result,
which is the one-loop
effective energy density for SU(2) in the presence of a
static chromomagnetic field with abelian background:
\begin{equation}
     {V(H)\over \Omega T}~
      =~{1\over 2}H^2~+~{11\over 48\pi^2}g^2H^2 \biggl(~
     \ln {gH\over \mu^2}~-~{1\over 2}~\biggr)
     ~+~\cdots    \ \ .
\end{equation}
The remarkable feature of this expression is that it
exhibits a minimum for $H$ different from zero, namely at
$ gH_{min}=\mu^2~\exp( -{24\pi^2 \over 11 g^2} ) $.
As Nielsen and Olesen \cite{niol} realized,
this minimum has unstable modes.
This instability can be seen from the fact
that there is an imaginary part of the effective energy density.
The imaginary part comes solely from the $n=0$ and
$S_z=-1$ contribution and in fact can be calculated directly,
\begin{equation}
    Im\biggl\lbrace ~{V(H)\over \Omega T}~\biggr\rbrace ~=
     ~-~ {c\over 2}~ Im\biggl\lbrace
    ~\int_{-\infty}^\infty dk~\sqrt{k^2-gH~} \biggr\rbrace
    = ~- {1 \over 8\pi}~g^2 H^2 .
\end{equation}
Note that the existence of the imaginary part is essential to obtain
asymptotic freedom. This is because in the absence of the
imaginary part the ultraviolet limit of
(\ref{garibaldi}) implies that the beta function will be the same as
the one of a scalar particle of mass $m^2=2gH$.
It is just the imaginary part which prevents us, after regulation,
from rotating the integration contour in the complex plane
and thereby spoiling the asymptotic freedom.

Although $H_{min}$ is not a classical solution there is a possibility
that non-perturbative effects might cause this configuration to
dominate the vacuum.
Since the above mentioned preliminary studies were done,
the property of this
non-trivial vacuum has been thoroughly investigated.
In particular a scenario, the so called
``Copenhagen vacuum'' \cite{niol} was proposed,
in which quantum fluctuations
and gluons condensations might create domains of constant
chromomagnetic configurations.

In reference \cite{mmrt}
this scenario was criticized arguing that in a strong
field configuration a perturbative analysis is unreliable,
and unstable configurations can be analyzed only by non-perturbative
methods. Therefore the
possible technique presently available to tackle this problem
is the lattice regularization. Monte Carlo simulation
can be used to generate the
typical vacuum configurations, and thus, to obtain direct information
about the vacuum structure.

\section{GENERAL RULES FOR IMPLEMENTING THE
            BACKGROUND FIELD METHOD ON LATTICE}

There are different ways to implement the background field method for
lattice gauge theory \cite{gross}.
The procedure presented here is quite general and has
the advantage that it can be applied with different kinds of sources.
Alternative methods are discussed
in \cite{fourD}.

Instead of the usual variable,
$ U^\mu_n~=~e^{ i a g \eta^\mu_b(n) T^b} $,
we introduce a new link variable in the
presence of a background field,
$ U({\cal A})^\mu_n=f(a,\eta^{({\cal A}) a}_\mu (n) )$,
in such a way that the continuum limit gives the expected
continuum expressions,
\begin{equation}
    \lim_{a \to 0}~\eta^{({\cal A}) b}_\mu (n)~
    =~\eta^b_\mu (x) ~+~ {\cal A}_\mu^b (x)
\end{equation}
\begin{equation}
    \lim_{ a \to 0}~\biggl( S[U({\cal A})^\mu_n]~
    -~S[U^\mu_n] \biggr)~
    =~S[\eta^a_\mu + {\cal A}_\mu^b]~-~S[\eta^a_\mu ],
\end{equation}
where $S$ is the usual action for lattice
gauge theories, and can eventually can be substituted by an
improved action.

The Euclidean generator functional is defined as:
\begin{equation}
     e^{-\tilde W[j,{\cal A}]} =
     { \int_{ bc(\eta^{({\cal A})} )} ~(dU)~
     e^{-S_{W}[U^{({\cal A})} ] ~
     -~j~\kappa[\eta{({\cal A})}] } \over
     \int_{bc(\eta)}~(dU)~ e^{-S_{W}[U]} } \ ,
\label{luria}
\end{equation}
where $\kappa$ is a functional of the source
that must be chosen to recover
the continuum limit in the scaling region.
The normalization here is chosen to be the same as the one without
the background field.
This choice is not the only possible one since the potential is
defined only up to an arbitrary additive constant.

The important point is that in the presence of a background
field we must take the boundary condition which makes
$\eta{({\cal A})}$ periodic,
$U({\cal A})^\mu_n=U({\cal A})^\mu_{n+La}$,
where L is the lattice size. However the normalization integral is
calculated with periodic boundary conditions for the links without
the constant chromomagnetic field, $U^\mu_n=U^\mu_{n+La}$.

Amongst the possible ways to define the link in the presence of
a background field, a natural choice is
\begin{equation}
    U({\cal A})^\mu_n~=~e^{ {i\over 2} a g  {\cal A}^b_\mu (n) T^b}
                 ~e^{ i a g \eta^b_\mu (n) T^b}
                 ~e^{ {i\over 2} a g {\cal A}^b_\mu (n) T^b} \ .
\label{montalban}
\end{equation}

We used three kinds of different sources in our computations.
For simplicity in the following,
we shall always choose the $z$ direction as the spatial direction of
the constant chromomagnetic field.
The first possibility is the so called
``abelian source'' because it points in the third color
direction. Such a source explores the Cartan subalgebra of SU(2),
and thus is called abelian. The natural counterpart is the
``diagonal source'', when the source is a combination of the generators
in the color directions $1$ and $2$. For simplicity we
choose the combination which is proportional to $T_\pm=T^1\pm iT^2$.
This source explores the non-diagonal matrix elements of SU(2).
A third possibility is the so called ``quadratic source''
where a combination which is parallel
to the external chromomagnetic field is used.

These three sources have advantages and disadvantages
when used in lattice Monte Carlo simulations.
We now describe the abelian source in more detail.
For the abelian source the term in equation (\ref{luria}) read as
\begin{equation}
     j~\kappa[\eta{({\cal A})}]~=
        ~\sum_{n,b,\mu}j~{a^4 \over 2}~\delta^{b3}~
        ( \delta_{\mu 2} - \delta{\mu 1} )~
        \eta^{({\cal A}) b}_\mu (n)  ,
\end{equation}
with $j$ as an arbitrary source strength.
Using the prescription given by (\ref{montalban}),
the link variable is
\begin{eqnarray}
     U({\cal A})^\mu_n~=&~U^\mu_n~~~~~~~~~~~~~(for~~ \mu~=~0,3)
                   \nonumber \\
     U({\cal A})^1_n~=&~e^{ - {i \over 4}  a^2 g H n_2 T^3}~U^1_n~
                 ~e^{ - {i \over 4}  a^2 g H n_2 T^3 }  \nonumber \\
     U({\cal A})^2_n~=&~e^{ {i \over 4} a^2 g H n_1 T^3 }~U^2_n~
                 ~e^{ {i \over 4}  a^2 g H n_1 T^3 } ,
\end{eqnarray}
n the presence of a background field (\ref{allen}).
Then equation (\ref{gadda}) becomes
\begin{equation}
     \tilde\Gamma[\tilde q,{\cal A}]~=
      ~\min_{ \{ j \} } ~\biggl[ \tilde
      W[j,{\cal A}]-(j,\tilde q) \biggr]  ,
\end{equation}
which can be obtained explicitly by (\ref{luria}).

\section{THE PRESENCE OF THE UNSTABLE MODE ON LATTICE }

We detect the presence of unstable
modes in the lattice is by analyzing the energy density.
Let us consider the contributions to the effective potential
from equation (\ref{garibaldi}).
As discussed above, the quantity,
\begin{equation}
    \sqrt{\nu_{k,n,s}~}~=~\sqrt{ gH~(2n+1+2S_z)~+~k_z^2~}
    \end{equation}
becomes imaginary for
$S_z=-1~$, $n=0$ and sufficiently small $k^2_z$ .
Due to the finiteness of the lattice $k_z$ is quantized as well,
as $k_z=2m \pi /L_z$, where $m$ is an integer.
The lowest inhomogeneous $z$-mode, $m=1$,
becomes stable for lattices whose
extent in the $z$-direction is smaller than
\begin{equation}
     L_z^{critical}~=~{2\pi \over \sqrt{gH} } \ .
\label{pertini}
\end{equation}
The homogeneous mode, $m=0$, which is always unstable,
is eliminated by imposing the condition
\begin{equation}
     \prod_{j=1}^{L_z}U^3_{z=j}=1 \ ,
\label{sppoly}
\end{equation}
in the path integral.
This enables us to search for the critical size,
$L_z^{critical}$, by
changing $H$ and looking for a sign of instability.

In these Monte Carlo simulations we generated a background field
$H$ in the $z$ direction and measured the expectation value of the
plaquette in the 1-2 plane $F_{12}$ as a
function of $\beta$ and $j$.
We used a heat bath updating procedure
with periodic boundary conditions
in presences of sources.
The special feature of this simulation was that
we forced the Polyakov line in the $z$ direction (\ref{sppoly})
to take a fixed value.
We made Monte Carlo simulations in a lattice volume
$L^3*L_z$, where $L$ is the size of the $x,y,t$ directions.
We bypassed the difficulty of needing for a huge lattice
because the directions $x$ , $y$ and $t$
do not have to be large.
In fact, they are irrelevant for the
homogeneous mode, m=0, and the dependence
on these directions shows up only through the excitation
of higher $n$-modes which
represent higher order perturbative effects.
Lattice artifacts might induce some small dependence
on $t,x,y$, but lattice simulations with
different $L_t*L_x*L_y$, but fixed $L_z$, show
insignificant dependence on these parameters,
as predicted theoretically.
We have varied $L_z$ from 4 to 50, and $L_t, L_x, L_y $ from
4 to 16 reaching a maximum lattice of $10^3*50$.
Unfortunately our programs were designed only to handle
even lattice sizes because of the checkerboard addresses.

We monitored the quantity
\begin{equation}
     X~=~{P_z[F_{12}(\beta,0)]~-~P_z[F_{12}(\beta,j)]
      \over j~P_z[F_{12}(\beta,0)]},
\label{wagner}
\end{equation}
where $P_z$ is the plaquette in the $z$ direction.
This quantity is proportional to the contribution
of the plaquette in the $z$ direction of
$\Delta E(F_{12}^{ext}/(F_{12}^{ext})^2$
and therefore is
very sensitive to the presence of the unstable modes.
We measured $X$ for different values of $\beta$
and $j$ performing, for large $j$ 4500 sweeps after discharging 500 for
thermalization. For smaller $j$ we increase the number of
sweeps until a significant amount of data was collected.
4 updating sweeps were made between subsequent measurements.

Our data show no sign of the unstable mode away from
the critical $\beta$ region ($\beta=2.1 - 2.5$).
The situation changes dramatically in the critical region where
the instability appears as a decrease of the vacuum
energy contribution to the plaquette in the $z$ direction.
This effect becomes more evident in the presence of
strong sources.

Thus far, we analyzed systematically
the critical region of $\beta$
from $2.10$ to $2.50$ with steps of $0.05$ for
$j=1.00, 0.75, 0.50, 0.25$; for some $\beta$ we also explored
$j=0.125 $ and $ j=0.1$.
We computed the value of $L_z^{critical}$
by interpolating for $X$ and taking the median value.
$L_z^{critical}$ was found to be dependent on the sources
strength $j$ and $\beta$ in this region.

To represent a physical quantity $L_z^{critical}$
must scale with $j$ according to equation (\ref{pertini}).
The manifestation of the rescaling is evident for
$\beta$  near $\beta=2.30$.
Moreover the agreement between the theoretical effective potential
predictions and the lattice simulation results is very good.

An important condition is that $L_z^{critical}$
should be greater than the deconfinement phase transition length,
i.e. $L_z^{critical}$ must belong to the confinement phase.
{}From the renormalization group considerations we have
\begin{equation}
     L_z^{critical}~\Lambda^{critical}~=~
     \bigl( {11 \over6\pi~\beta} \bigr)^{-51/121}
     ~e^{ {-3\pi^2\beta \over 11 } }~
     \biggl[ 1~+~O\biggl({1\over \beta}\biggr)\biggr]  .
\end{equation}
Hence the ratio between
the $\Lambda^{critical}$ and the renormalization scale
parameter $\Lambda$
is independent of $\beta$.
Using the well known value of $\Lambda$ given by
\cite{kpsk}, we have
\begin{eqnarray}
        {\Lambda \over \Lambda^{critical}(j=1.00) }~=
         &~ 1.226 \pm 0.019         \nonumber \\
        {\Lambda \over \Lambda^{critical}(j=0.75)}~=
           &~ 1.390 \pm 0.025       \nonumber \\
        {\Lambda \over \Lambda^{critical}(j=0.50)}~=
           &~ 1.572 \pm 0.026       \nonumber \\
        {\Lambda \over \Lambda^{critical}(j=0.25)}~=
           &~ 1.934 \pm 0.031  .
\label{diderot}
\end{eqnarray}
It is clear from these results that $L_z^{critical}$
belongs to the confinement phase and
there is a good agreement with the renormalization
group equation (see Fig. 2).

Using our data we also evaluated
$L_{deconf}$ as a function of $\beta$.
These values are obtained by comparing
the plaquette in the $t$ and $z$ directions.
{}From $L_{deconf}$ using the renormalization equation
we estimated
the renormalization scale parameter $\Lambda$.
This data are compatible with
reference \cite{kpsk} where simulations were done
specifically to analyze the deconfinement transition.
Nevertheless the comparison is interesting
because it shows
that from our data we can extract $L_{deconf}$
which is clearly distinct from $L_z^{critical}(j)$.

The limit $j \rightarrow 0$ is obtained by studying
the ratio (\ref{diderot}) for several $j$ and then
extrapolating to zero.
The result is
$\Lambda / \Lambda^{critical}(j=0) = 2.5  \pm  0.2 $,
which corresponds to
$L_z^{critical}(\beta=2.3, j=0) ~\sim ~ 18 $.
The ratio of the characteristic lengths is found to be
\begin{equation}
       {L_z^{critical}\over L_{deconf}} ~=~ 2.5 ~\pm 0.2.
\label{massanet}
\end{equation}
It is worthwhile noting that the energy density
measured by (\ref{wagner})
shows a small peak at
$\tilde L_z\approx 2 L_z^{critical}$,
indicating the onset of the instability of the next
$k=4\pi/L_z$, mode. One may expect
weaker singularities at $\approx m L_z^{critical}$
which correspond to higher momentum modes, too.

\section{SUMMARY AND CONCLUSIONS}

We support the analysis made by \cite{mmrt}
showing that the vacuum structure of SU(2)
is a non-perturbative effect, necessitating
lattice regularization. In particular we analyzed
the difference between stable and unstable
configurations, and the origin of the instability
for SU(2).

Near the critical value of the coupling
constant (near $\beta = 2.3$)
we detected the presence of the unstable mode
for the first time by Monte Carlo simulations.
In particular by changing the lattice volume and
the source $j$ we are able to turn the unstable mode on and off .
This allowed us to analyze the behavior of the unstable mode,
showing that it has the correct behavior
in the limit $j \rightarrow 0$ and
under the renormalization group equations.

We found that there is a new length scale,
$L^{critical}$, in the theory
given by (\ref{massanet})
which is significantly larger than
the confinement radius, $L_{deconf}$.
This is rather puzzling since no
correlations were expected to show up
beyond the confinement radius, $L_{deconf}$.
Another surprising
result was the appearance of the
singularity driven by the instability of the
higher momentum mode with momentum $k=4\pi/L_z$.
This suggests the presence of "resonances"
corresponding to even longer length scales,
$mL^{critical}$.

One may interpret our results by exchanging
the z and the time directions.
In this language we found singularities
in the pressure at temperatures
below the deconfinement transition
driven by unstable electric condensate.
In order to make more precise conclusions
one has to  delve in to
the continuum and the $j\to0$ limit.

We can establish two conditions that
the lattice system should satisfy in order to
reflect the richness of the
non-perturbative vacuum of the continuum theory.
One such condition is that a lattice size greater than
$L_z(\beta, j=0)$ is needed to display the instability.
Another one comes from the
observation that the instability disappears
outside the scaling window.
Thus this instability is a
distinguishing feature of the continuum
rather than the strong coupling vacuum.
In order to study the continuum
theory we should remain in the region
where the instability is manifest.
The disappearance of the instability
as $\beta\to0$ may provide a clue to
understanding the difference between
the physics of the continuum and the
strong coupling lattice theory.

\acknowledgements

It is a pleasure to thank
Ken Johnson and Suzhou Huang
for several useful discussions.

\vfill
\eject
\centerline{\bf FIGURE CAPTIONS}

Figure 1.

The quantity
$X={P_z(\beta,0)-P_z(\beta,j)\over jP_z(\beta,0) }$
as a function   of the lattice size $L_z$ for
$\beta=2.35$ and $j=1.0$ (stars); $j=0.75$ (squares);
$j=0.50$ (crosses); $j=0.25$ (diamonds).

\bigskip

Figure 2.

$L_z^{critical}(j=1.00)$ (stars),
$L_z^{critical}(j=0.75)$ (squares),
$L_z^{critical}(j=0.50)$ (crosses),
$L_z^{critical}(j=0.25)$ (diamonds),
with their renormalization group
dependence fit (dashed lines);
and the deconfinement transition from
reference \cite{kpsk} (full line).


\begin{references}

\bibitem{paris}  A.R. Levi, ``Subtleties and Fancies in Gauge Theory non
                 Trivial Vacuum'', Proceedings of the ``Quantum
                 Infrared Physics Workshop'', (World Scientific 1985).

\bibitem{fourD} J. Ambj{\o}rn {\it et al.},
                Phys. Lett. {\bf 225B} (1989) 153;
                {\bf 245B} (1990) 575;
                P. Cea and L. Cosmai,
                Phys. Rev. {\bf D43} (1991) 620;
                Phys. Lett. {\bf 264B} (1991) 415;
                A.R. Levi, Nucl. Phys. (proc. supl.)
                {\bf B34} (1994) 161.

\bibitem{threeD} H.D. Trottier, Phys. Rev. {\bf D44} (1991) 464;
                 H.D. Trottier and R.M. Woloshyn,
                 Phys. Rev. Lett. {\bf 70} (1993)  2053;
                 P. Cea and L. Cosmai, Phys. Rev. {\bf D48} (1993) 3364.

\bibitem{dewi} B.S. DeWitt, Phys. Rev. Lett. {\bf 162} (1967) 1195;
               G. 't Hooft, Nucl. Phys. {\bf B62} (1973) 444;
               L.F. Abbott, Nucl. Phys. {\bf B185} (1981) 189.

\bibitem{lehu} S. Huang and A.R. Levi, Phys. Rev.
               {\bf D49} (1994) 6849.

\bibitem{const} L.S. Brown and W.I. Weisberger,
                Nucl. Phys. {\bf B157} (1979) 285.

\bibitem{forgot} J. Ambj{\o}rn, N.K. Nielsen and P. Olesen,
                 Nucl. Phys. {\bf B152} (1979) 75.

\bibitem{mmrt} L. Maiani, G. Martinelli, G.C. Rossi and M. Testa,
                Nucl. Phys. {\bf B273} (1986) 275.

\bibitem{niol}  N.K. Nielsen and P. Olesen,
                Nucl. Phys. {\bf B144} (1978) 376;
                Phys. Lett. {\bf 79B} (1978) 304;
                H.B. Nielsen and M. Ninomiya ,
                Nucl. Phys. {\bf B156} (1979) 1;
                J. Ambj{\o}rn and P. Olesen, Nucl. Phys.
                {\bf B170} (1980) 60 and 265;
                J. Ambj{\o}rn {\it et al.},
                Nucl. Phys. {\bf B175} (1980) 349;
                R. Parthasarathy {\it et al.},
                Can. Jour. Phys  {\bf 61} (1983) 1442;
                R. Anishetty {\it et al.},
                J. Phys  {\bf G16} (1990) 375 and 1187.

\bibitem{savv} G.K. Savvidy, Phys. Lett. {\bf 71B} (1977) 133;
               S.G. Matinyan and G.K. Savvidy,
               Nucl. Phys. {\bf B134} (1978) 539.

\bibitem{gross} R. Dashen and D.J. Gross
                Phys. Rev. {\bf D23} (1981) 2340.

\bibitem{kpsk} J. Kuti, J. Polonyi and K. Szlachanyi
               Phys. Lett. {\bf 98B} (1981) 199;
               E. Kovacs, Phys. Lett. {\bf 118B} (1982) 125.




\end{references}
\end{document}